\documentclass[aps,pra,onecolumn,superscriptaddress]{revtex4-2}

\usepackage{graphicx}
\usepackage{dcolumn}
\usepackage{bm}
\usepackage{amsmath}
\usepackage{amssymb}
\usepackage{latexsym}
\usepackage{epsfig}
\usepackage{amsbsy}
\usepackage{array}
\usepackage{amssymb}
\usepackage{setspace}
\usepackage{bm}
\usepackage{epstopdf}
\usepackage{subfigure}
\usepackage{color}
\usepackage{siunitx}
 \usepackage{tabularray}
\usepackage{multirow}

\def\sint{\ifmmode{- \!\!\!\!\!\! \int}
    \else{\hbox{$- \!\!\!\! \int \ $}}\fi}

\begin{document}
\title{Fluctuation-induced quenching of chaos in quantum optics}
\author{Mei-Qi Gao}
\affiliation{College of Sciences, Northeastern University, Shenyang 110819, China}
\author{Song-hai Li}
\affiliation{College of Sciences, Northeastern University, Shenyang 110819, China}
\author{Xun Li}
\affiliation{National Key Laboratory of Shock Wave and Detonation Physics, Institute of Fluid Physics, China Academy of Engineering Physics, Mianyang 621900, China}
\author{Xingli Li}
\affiliation{Department of Physics, The Chinese University of Hong Kong, Shatin, New Territories, Hong Kong, China}
\affiliation{Lanzhou Center for Theoretical Physics, Key Laboratory of Theoretical Physics of Gansu Province, Key Laboratory of Quantum Theory and Applications of MoE, Gansu Provincial Research Center for Basic Disciplines of Quantum Physics, Lanzhou University, Lanzhou 730000, China}
\author{Jiong Cheng}
\affiliation{Department of Physics, School of Physical Science and Technology, Ningbo University, Ningbo 315211, China}
\author{Wenlin Li}
\email{liwenlin@mail.neu.edu.cn}
\affiliation{College of Sciences, Northeastern University, Shenyang 110819, China}

\date{\today}
\begin{abstract}
Recent studies have extensively explored chaotic dynamics in quantum optical systems through the mean-field approximation, which corresponds to an ideal, fluctuation-free scenario. However, the inherent sensitivity of chaos to initial conditions implies that even minute fluctuations can be amplified, thereby questioning the applicability of this approximation. Here, we analyze these chaotic effects using stochastic Langevin equations or the Lindblad master equation. For systems operating at frequencies of $10^5$ to $10^7$ Hz, we demonstrate that room-temperature thermal fluctuations are sufficient to suppress chaos at the level of expectation values, even under weak nonlinearity. Furthermore, nonlinearity induces deviations from Gaussian phase-space distributions of the quantum state, revealing attractor-like features in the Wigner function. With increasing nonlinearity, the noise threshold for chaos suppression decreases, approaching the scale of vacuum fluctuations. These results provide a bidirectional validation of the quantum mechanical suppression of chaos.
\end{abstract}
\maketitle
\section{Introduction}  
Chaotic dynamics, as a quintessential nonlinear phenomenon, has profoundly influenced fields beyond fundamental physics~\cite{book1,Abarbane1993}, with applications in secure communications~\cite{Lu2016,Ermann2016}, remote sensing~\cite{Hu2003}, finance, and sociology~\cite{Hsieh1991}. In recent years, the rapid advances in quantum optics and quantum information science have spurred interdisciplinary investigations at the intersection of nonlinear dynamics and quantum optical systems. For instance, phenomena such as Hopf bifurcations~\cite{Piergentili2021}, self-sustained oscillations~\cite{Marquardt2006,Li2023}, and spontaneous synchronization~\cite{Holmes2012,Heinrich2011,Mari2013} have been theoretically analyzed and experimentally observed in mesoscopic quantum platforms, such as optomechanical and magnetomechanical devices~\cite{Bagheri2013,Sheng2020,Cheng2023}, thereby enabling novel quantum information processing protocols~\cite{Li2017}.

Originally defined within classical mechanics, these phenomena are frequently modeled in quantum contexts via mean-field approximations, which simplify the treatment of nonlinear dynamics~\cite{Heinrich2011,Bemani2017,Shi2026,Peng2024,Ghosh2026,Yang2019,Saikoa2025,Lu2015,Larson2011,Wu2025,Sun2025,Xu2025,Zhang2010,Xu2024}. However, fully quantum simulations of limit-cycle systems have recently demonstrated that, even under strong driving and weak nonlinearity, the absence of dissipation in the phase direction causes the oscillator state to depart from Gaussian statistics, forming ring-like structures in phase space called a limit cycle state~\cite{Lee2013,Weiss2015,Li2020,Li2023}. Concurrently, the expectation values of the associated observables cease to evolve temporally, attaining asymptotic steady states. This insight aligns with developments in time-crystal research, which, from the vantage of time-translation symmetry preservation, establishes a no-go theorem prohibiting spontaneous quantum oscillatory dynamics~\cite{Iemini2018,Watanabe12015}.

Intuitively, chaos exhibits stronger nonlinear traits than limit-cycle dynamics, rendering the mean-field approximation more susceptible to breakdown. This fragility stems from a hallmark of chaos: the exponential amplification of minute initial fluctuations by the nonlinear dynamics. Here, we exemplify these effects using a paradigmatic optomechanical system, a common platform in recent studies of quantum chaos. We first revisit its dynamics via the stochastic Langevin equations~\cite{Lee2013,Li2020,Li2023}, revealing that the apparent persistent irregularity in the expectation values dissipates over time, yielding a monotonic approach to an asymptotically stable state that upholds time-translation symmetry. Phase-space analysis of the steady state further discloses a well correspondence between the chaotic distribution and the mean-field trajectory, akin to the limit-cycle regime where phase undergoes diffusive spreading along the attractor. Finally, full quantum simulations, performed via the quantum Lindblad master equation in a truncated Hilbert space, corroborate these semiclassical findings; notably, the intrinsic noise now arises from irreducible vacuum fluctuations dictated by the Heisenberg uncertainty principle, rather than coolable thermal contributions. These results substantiate the quantum mechanical suppression of chaos~\cite{Bakemeier2015}.

This paper is organized as follows. In Sec.~\ref{System dynamics}, we present the mean-field approximation equations and the stochastic Langevin equations for the system. In Sec.~\ref{The influence of noise on chaotic dynamics}, we analyze the influence of thermal noise on the evolution of the system's expectation values and the corresponding phase-space distributions beyond the mean-field approximation. The full quantum analysis is presented in Sec.~\ref{full quantum analysis}. Finally, in Sec.~\ref{Discussion and conclusion}, we summarize our findings and discuss similar conclusions for other types of systems.
\section{System dynamics}  
\label{System dynamics}
We consider a standard optomechanical system comprising a cavity and an oscillator, as illustrated in Fig.~\ref{fig:1}. In addition to the nonlinearity arising from radiation pressure, we incorporate a Kerr medium within the cavity to enhance the nonlinearity and to more readily elucidate chaotic dynamics~\cite{Khan2015,Li2025,Sun2024}. A driving field of amplitude $E$ and frequency $\omega_d$ is injected into the cavity from the side of the fixed mirror. The total Hamiltonian of the system is then given by~\cite{Aspelmeyer2014,Sun2024}:
\begin{figure}[]
\centering
\includegraphics[width=3.5in]{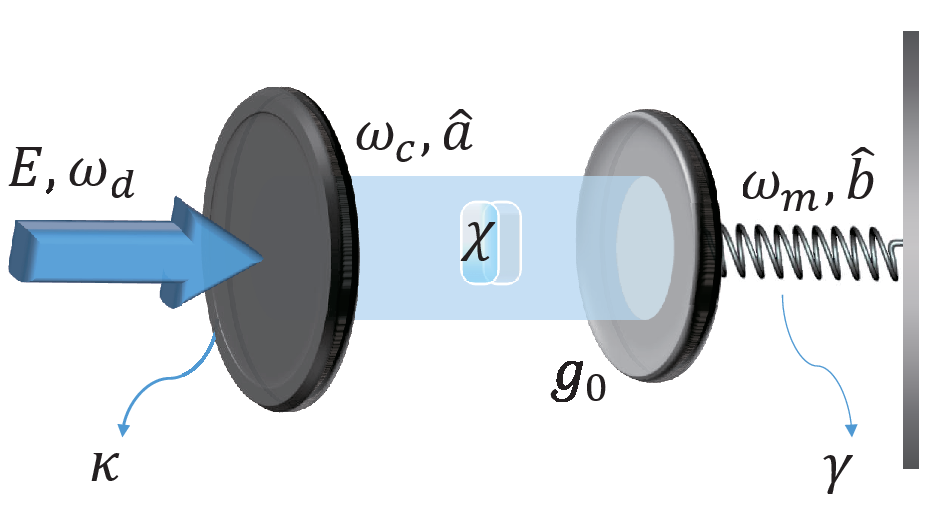}
\caption{Schematic of the optomechanical system, consisting of a Fabry-P\'erot cavity with one movable end mirror coupled to a mechanical oscillator. The cavity is filled with a Kerr nonlinear medium.
\label{fig:1} }
\end{figure}
\begin{equation}
\begin{split}
\hat{H}/\hbar=&\omega_c\hat{a}^\dagger\hat{a}+\omega_b\hat{b}^\dagger\hat{b}-g\hat{a}^\dagger\hat{a}\left(\hat{b}^\dagger+\hat{b}\right)+iE(\hat{a}^\dagger e^{-i\omega_d t}-\hat{a}e^{i\omega_d t})-\chi\left(\hat{a}^\dagger\hat{a}\right)^2,
\end{split}
\label{eq:total_H}
\end{equation}
where $\hat{a}$ ($\hat{a}^\dagger$) and $\hat{b}$ ($\hat{b}^\dagger$) are the annihilation (creation) operators for the cavity mode and mechanical oscillator, respectively; $\omega_c$ and $\omega_b$ are the corresponding resonant frequencies; $g$ is the optomechanical coupling rate; and $\chi$ is the Kerr nonlinearity coefficient. $E$ describes the driving strength of the laser field, which can be expressed as $E =\sqrt{2\kappa_{\rm in}P/\hbar\omega_d}$. Here $P$ is the input power, and $\kappa_{\rm in}$ is the dissipation rate of the fixed mirror. Following the procedures of our previous work, we derive the Heisenberg-Langevin equations in the Heisenberg picture by including the coupling to the thermal baths. In a frame rotating at the drive frequency $\omega_d$, these equations read~\cite{book2}:
\begin{equation}
\begin{split}
\dot{\hat{a}}=&\left\{i\left[\Delta+g\left(\hat{b}^\dagger+\hat{b}\right)+\chi\left(\hat{a}^\dagger\hat{a}+\hat{a}\hat{a}^\dagger\right)\right]-\kappa\right\}\hat{a}+E+\sqrt{2\kappa}\hat{a}_{\rm in}, \\
\dot{\hat{b}}=&(-i\omega_b-\gamma)\hat{b}+ig\hat{a}^\dagger\hat{a}+\sqrt{2\gamma}\hat{b}_{\rm in}.
\end{split}
\label{eq:qle}
\end{equation}
Here, $\kappa$ and $\gamma$ are the dissipation rates of the cavity field and mechanical resonator, respectively. The input noise operators $\hat{a}_{\rm in}$ and $\hat{b}_{\rm in}$ satisfy the properties of Gaussian white noise, i.e., they have zero expectation value and their only nonvanishing second-order correlation is $\langle \hat{o}_{\rm in}^\dagger(t)\hat{o}_{\rm in}(t')+\hat{o}_{\rm in}(t')\hat{o}^\dagger_{\rm in}(t)\rangle=(2\bar{n}_o+1)\delta(t-t')$, for $o\in\{a,b\}$, where $n_o= [\exp(\hbar\omega_o/k_bT)-1]^{-1}$ is the mean thermal excitation number of the corresponding mode~\cite{Giovannetti2001}.
 
By taking the expectation value of both sides of Eq.~(\ref{eq:qle}) and neglecting all correlations (i.e., employing the mean-field approximation as $\langle \hat{o}_1\hat{o}_2\rangle\simeq \langle \hat{o}_1\rangle\langle\hat{o}_2\rangle$ where $\hat{o}_j$  is an arbitrary operator in the Langevin equation~\cite{Vitali2007}), we obtain a set of self-consistent equations for the expectation values:
\begin{equation}
\begin{split}
\langle \dot{\hat{a}}\rangle=&\left\{i\left[\Delta+g\left(\langle\hat{b}\rangle^*+\langle\hat{b}\rangle\right)+2\chi\vert\hat{a}\vert^2\right]-\kappa\right\}\vert\langle\hat{a}\rangle\vert+E, \\
\langle\dot{\hat{b}}\rangle=&(-i\omega_b-\gamma)\langle\hat{b}\rangle+ig\vert\langle\hat{a}\rangle\vert^2.
\end{split}
\label{mean_value}
\end{equation}
Equation~\eqref{mean_value} provides us with information about the mean-field approximation discussed below. It is noteworthy that, even if we adopt an alternative perspective on the mean-field approximation, that is, expressing the operator as the sum of its expectation value and fluctuations ($\hat{o} = \langle \hat{o} \rangle + \delta o$)~\cite{Xu2024}, the above equation remains an approximation. This arises because, from a logical standpoint, decoupling an equation that incorporates both the expectation value, a $c$-number, and fluctuation operators into separate independent equations is rigorous only if we neglect $\langle \delta o_1 \delta o_2 \rangle$ which has a non-zero value. In fact, one must treat $\delta o_1 \delta o_2=\langle \delta o_1 \delta o_2 \rangle+
\delta (\delta o_1 \delta o_2)$ as a new operator, incorporate its expectation value into Eq.~\eqref{mean_value}, and then disregard its fluctuation term.
\begin{figure}[]
\centering
\includegraphics[width=4in]{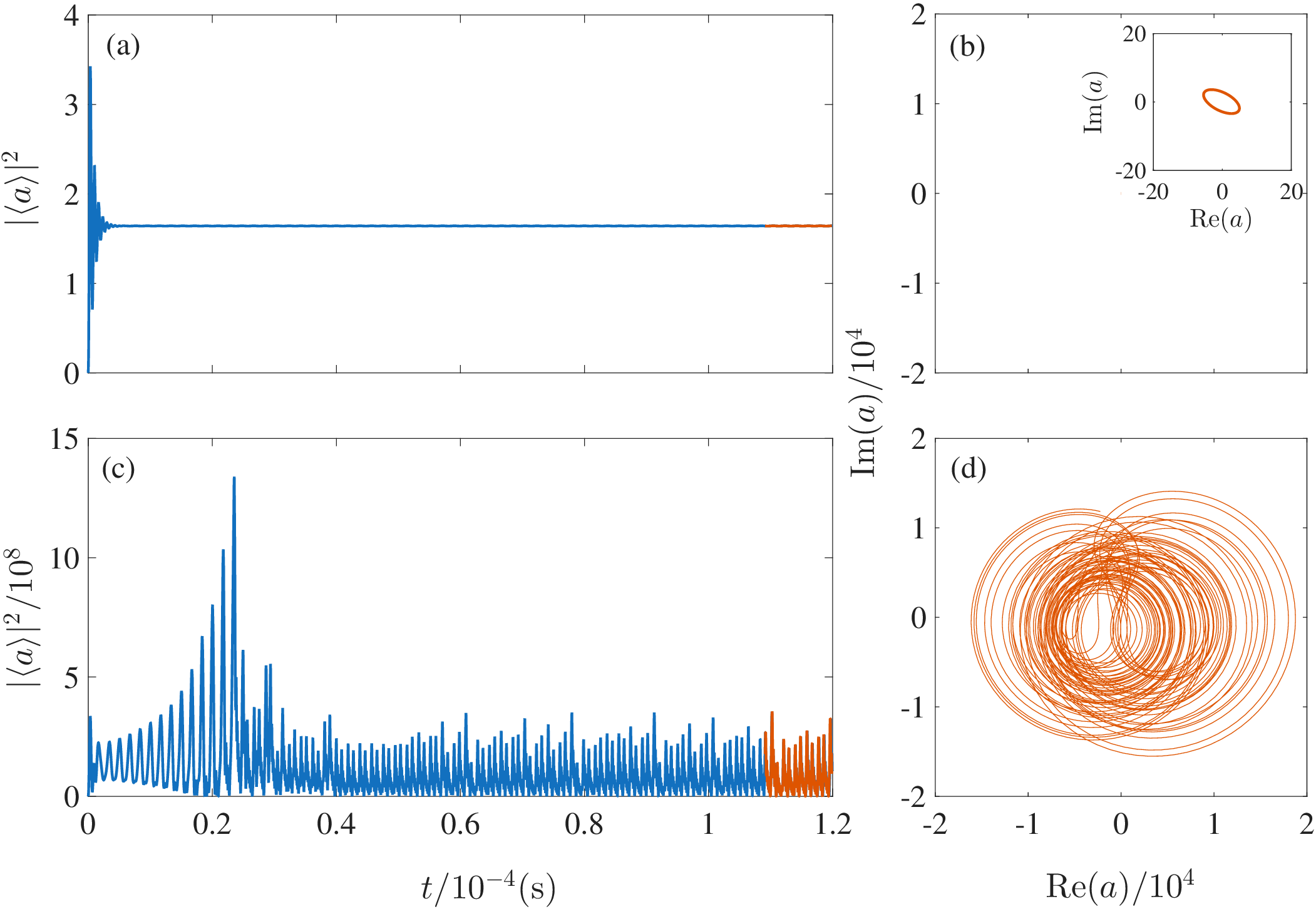}
\caption{(a),(c): Time evolution of the intracavity light intensity obtained from the mean-field Eq.~\eqref{mean_value} for the weakly nonlinear ($g/2\pi = 1$\,Hz) and strongly nonlinear regimes $g/2\pi = 25$\,Hz), respectively. (b),(d) Phase-space projections of the trajectory segments highlighted in orange in (a) and (c). The parameters used for this simulation are $\omega_m/2\pi = 525$\,kHz, $Q = \omega_m/\gamma = 10^{7}$, $P = 0.4$\,mW, $\chi/2\pi = 1.625 \times 10^{-3}$\,Hz, $\kappa/2\pi = 220$\,kHz, $\kappa_{\rm in}/2\pi = 100$\,kHz, and $\omega_d/2\pi = 10^{14}$\,Hz. The corresponding dimensionless parameter are: $\gamma=10^{-7}$, $\chi \simeq 3.095 \times 10^{-9}$, $\kappa \simeq 0.4190$, $E\simeq 26404$ by setting $\omega_m=1$ as the unit.
\label{fig:2}}
\end{figure}

The mean-field approximation appears justified, as the fluctuation terms are initially negligible under conditions of weak nonlinearity and strong driving. However, Ludwig~\textit{et al.}~\cite{Ludwig2013}, Weiss~\textit{et al.}~\cite{Weiss2015} and Navarrete-Benlloch~\textit{et al.}~\cite{Navarrete-Benlloch2017} recognized that these small terms can accumulate over time. In the absence of a dissipation mechanism, they continue to build up throughout the system's evolution, ultimately exerting a substantial influence on the dynamics. A paradigmatic example is a self-sustained limit-cycle oscillator, in which the dynamics exhibit no dissipation along the phase direction. Consequently, phase fluctuations accumulate indefinitely, leading to diffusion of the entire state along the classical trajectory in phase space and thereby forming an annular distribution~\cite{Lee2013,Navarrete-Benlloch2017,Li2021}. We consider scenarios in which the Wigner function of the initial state is non-negative and the nonlinearity, while sufficient to drive the state away from Gaussianity, is too weak to generate Wigner negativity~\cite{book2,Lee2013,Li2021}. Under these conditions, the quantum Langevin equation is accurately approximated by averaging over multiple trajectories governed by the following stochastic equations for $c$-numbers~\cite{Li2017,Li2020,Rodrigues2010}:
\begin{equation}
\begin{split}
\dot{a}=&\left\{i\left[\Delta+2g\text{Re}\left(b\right)+2\chi\left(\vert a\vert^2\right)\right]-\kappa\right\}{a}+E+\sqrt{2\kappa}{a}_{\rm in}, \\
\dot{{b}}=&(-i\omega_b-\gamma){b}+ig\left(\vert a\vert^2-\dfrac{1}{2}\right)+\sqrt{2\gamma}{b}_{\rm in}.
\end{split}
\label{eq:cle}
\end{equation}
The noise terms for these $c$-numbers satisfy $\langle o_{\rm in}^*(t)o_{\rm in}(t')\rangle=(\bar{n}_o+1/2)\delta(t-t')$, reflecting the loss of commutation relations~\cite{book2,Li2020}. Indeed, as long as the system contains significantly more than a few quanta, the stochastic and quantum Langevin equations yield consistent results~\cite{Lee2013,Li2023}. By performing $N$ independent simulations of Eq.~\eqref{eq:cle}, expectation values and variances are obtained as $\langle \hat{o} \rangle = N^{-1} \sum_{j=1}^N o^j$ and $\langle (\delta \hat{o})^2 \rangle = N^{-1} \sum_{j=1}^N (o^j)^2 - (N^{-1} \sum_{j=1}^N o^j)^2$, respectively, while correlations are computed via $\langle \delta \hat{o}_1 \delta \hat{o}_2 \rangle = N^{-1} \sum_{j=1}^N o_1^j o_2^j - (N^{-1} \sum_{j=1}^N o_1^j)(N^{-1} \sum_{j=1}^N o_2^j)$, where the superscript $j$ denotes the result of the $j$th trajectory. In this representation, the state of the system can also be characterized by the Wigner function, which, in the non-negative regime, corresponds to a classical phase-space probability distribution. The latter is constructed from the $c$-number outcomes as~\cite{Li2020}:
\begin{equation}
\begin{split}
W[\text{Re}(o),\text{Im}(o)]=\lim_{h\rightarrow 0}\dfrac{N_{\text{Re}(o),\text{Im}(o)}}{h^2},
\end{split}
\label{eq:cle_W}
\end{equation}
where $N_{\text{Re}(a),\text{Im}(a)}$ denotes the number of outcomes satisfying $\text{Re}(o^j)\in(\text{Re}(o^j)-h/2,\text{Re}(o^j)+h/2]$ and $\text{Im}(o^j)\in(\text{Im}(o^j)-h/2,\text{Im}(o^j)+h/2]$. Here, $h$ is the interval in phase space.

\section{The influence of noise on chaotic dynamics}  
\label{The influence of noise on chaotic dynamics}
\begin{figure}[]
\centering
\includegraphics[width=4in]{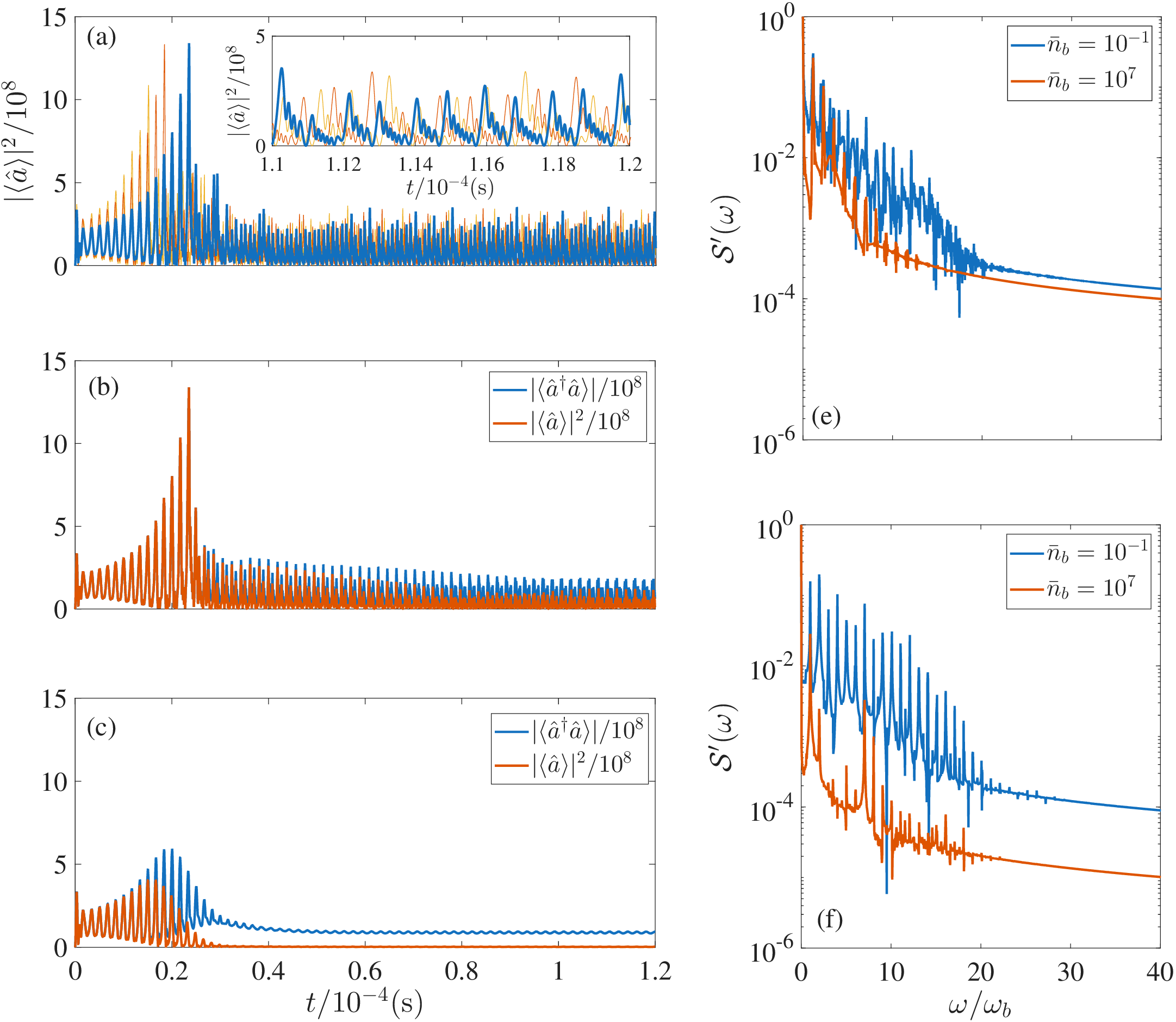}
\caption{ (a): Comparison between the mean-field solution and individual stochastic trajectories. The blue curve shows the intracavity light intensity obtained from the mean-field equation~\eqref{mean_value}, while the red and yellow curves depict two trajectories obtained from the stochastic Langevin equation~\eqref{eq:cle} with $n_b = 10^{7}$. The inset is the magnified view of the late-time evolution. (b) and (c): Intracavity light intensity obtained from $250,000$ realizations of Eq.~\eqref{eq:cle} for $n_b = 10^{-1}$ and $n_b = 10^{7}$, respectively. The Fourier transform of the red curves in (b) and (c) is plotted in (e) corresponding to the first half of the evolution ($t=0$ to $0.6\times 10^{-4}$~s) and (f) correspondin to the second half ($0.6\times 10^{-4}$ to $1.2\times 10^{-4}$~s). All other parameters are as in Fig.~\ref{fig:2}.
\label{fig:3}}
\end{figure}

In this section, the dynamical results are compared. As shown in Figs.~\ref{fig:2}(a) and (b), the system exhibits increasingly irregular dynamics as the nonlinearity increases, consistent with expectations. The final segment of the time-domain trajectory (highlighted in orange) is projected into phase space, where it displays the characteristics of a chaotic attractor.
When fluctuation and noise are included in the simulations, as shown in Fig.~\ref{fig:3}(a), individual trajectories remain irregular. However, comparison of noisy trajectories with the mean-field solution reveals temporal shifts in the peak positions induced by noise. This behavior exemplifies the hallmark of chaos, in which small initial perturbations and noise are amplified during the evolution. These results suggest that ensemble averaging over multiple trajectories may suppress the irregular features. As shown in Figs.~\ref{fig:3}(b) and (c), the ensemble average of $\hat{a}$ computed from many trajectories indeed exhibits regular dynamics, with the irregular oscillations effectively averaged out. We refer to this phenomenon as the annealing of chaos. In particular, Fig.~\ref{fig:3}(b) corresponds to $n_b=0.1$, where the dominant noise arises from zero-point fluctuations associated with the Heisenberg uncertainty principle. In this case, irregularity persists to some degree, though it is weaker than in the mean-field prediction. Figure~\ref{fig:3}(c) corresponds to the room-temperature case ($n_b\sim 10^{7}$), where the irregular dynamics are almost completely suppressed. To illustrate the magnitude of the fluctuations, $\vert\langle \hat{a}^\dagger\hat{a} \rangle\vert$ and $\vert\langle \hat{a} \rangle\vert^2$ are plotted separately. As evident in Fig.~\ref{fig:3}(c), larger environmental phonon numbers yield fluctuations whose amplitude substantially exceeds that of the mean value. 
This evolution from irregular to regular behavior is more readily discerned in the frequency domain. To illustrate this, we compute the Fourier transform of the light intensity $\langle \hat{a}^\dagger\hat{a}\rangle$ obtained from the stochastic Langevin equation, that is,
\begin{equation}
\begin{split}
\mathcal{S}(\omega)=\left\vert \dfrac{1}{\sqrt{2}}\int dt \langle \hat{a}^\dagger\hat{a}\rangle(t)e^{-i\omega t}\right\vert,
\end{split}
\label{eq:FFT}
\end{equation}
and present the corresponding normalized spectra $\mathcal{S}'=\mathcal{S}/\max(\mathcal{S})$ in Figs.~\ref{fig:3}(e) and (f). The spectrum in Figs.~\ref{fig:3}(e) corresponds to the initial time interval ($t=0$s to $0.6\times 10^{-4}$s), capturing the transient dynamics of the system. In contrast, the spectrum in Figs.~\ref{fig:3}(f) pertains to the subsequent interval ($0.6\times 10^{-4}$s to $1.2\times 10^{-4}$s), reflecting the regime in which the system undergoes attractor evolution. As evident from the spectra, when the corresponding thermal excitation number of the reservoir is large, the higher-order spectral components are markedly suppressed. Relative to the low-temperature environment case, this suppression reduces the higher-order contributions by at least two orders of magnitude. In particular, in Fig.~\ref{fig:3}(f), the blue curve retains the signatures of higher-order sidebands, whereas the red curve effectively corresponds to a single-mode oscillation.

\begin{figure}[]
\centering
\includegraphics[width=4in]{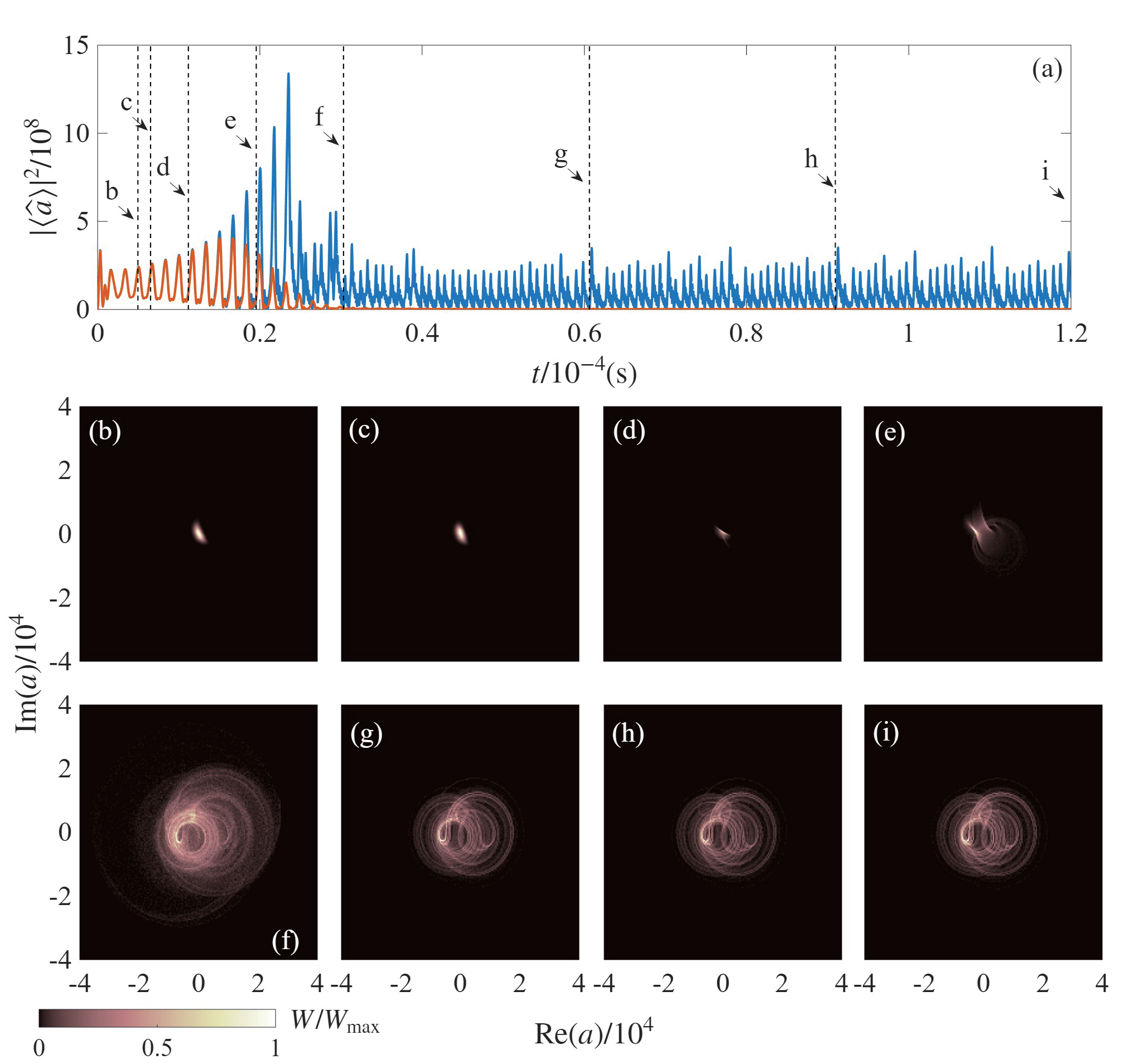}
\caption{ The blue line reproduces the mean-field light intensity from Fig.~3(a), while the red line shows the averaged light intensity from $250,000$ stochastic trajectories with $n_b=10^7$ [cf.~Fig.~3(c)]. Eight representative time points, labeled (b)--(i), are selected across the dynamical evolution; for each, the normalized probability distribution of the optical field $a$ is displayed in the corresponding subpanel. All other parameters are as in Fig.~\ref{fig:3}.
\label{fig:4}}
\end{figure}

For convenience and reference in the discussion below, Fig.~\ref{fig:4}(a) shows the light intensity $\vert \langle\hat{a}\rangle\vert^2$ obtained by mean-field equations~[blue line in Fig.~\ref{fig:2}(c)] together with that obtained from the stochastic Langevin equations for $n_b=10^7$~[red line in Fig.~\ref{fig:3}(c)]. As shown in Figs.~\ref{fig:4}(b)-(i), the outcomes of $250,000$ simulations at different times are plotted as probability distribution functions in phase space. These probability distribution functions represent the system state at a fixed instant, with the corresponding time for each phase distribution marked in panel~(a). Initially, the system remains in a Gaussian state, during which the evolution of the expectation value from the stochastic Langevin equation closely matches that of the mean-field equation. Subsequently, the system deviates from the Gaussian state, rendering the evolution of the expectation value more regular. The distribution function, or equivalently the scatter plot of the reconstructed distribution, aligns well with the phase-space trajectory obtained from the mean-field approximation. Indeed, these distributions can be interpreted as adhering to the dynamics predicted by the mean-field theory. However, variations in the phase or temporal degrees of freedom, wherein distinct simulation outcomes correspond to different instants along the mean-field trajectory, ultimately result in the probability distribution of the system's state in phase space matching the spread of the mean-field trajectory.

This implies that the system density operator is effectively a statistical mixture of states each exhibiting chaotic trajectories,
\begin{equation}
\begin{split}
\rho = \sum_i P_i \rho_i^c,
\end{split}
\label{eq:rho}
\end{equation}
where $\rho_i^c$ denotes a state with a chaotic trajectory, while the overall density operator $\rho$ exhibits asymptotically steady-state behavior~\cite{Navarrete-Benlloch2017,Li2021}. Notably, alternative decompositions always exist; for instance, $\rho = \sum_i P'_i \rho_i^{\prime c}$, where the $\rho_i^{\prime c}$ may be chosen as states with regular trajectories. No measurement can distinguish between these decompositions. Consequently, neither the individual $\rho_i^c$ nor $\rho_i^{\prime c}$ terms possess observable or physical significance. Thus, the apparent chaotic state during annealing is not intrinsically chaotic; equivalently, the chaos predicted by mean-field theory is unphysical.

It is also worth noting that, under continuous weak zero-difference or heterodyne measurements from the outset, the quantum Zeno effect results in the system being described by a stochastic pure state (rather than a density matrix), governed by the stochastic Schr\"odinger equation~\cite{Es’haqi-Sani2020}. In the classical limit, individual stochastic trajectories exhibit chaotic evolution~\cite{Verlot2009}. In this scenario, the components $\rho_i^c$ and $\rho_i^{\prime c}$ become distinguishable, or, more precisely, the $\rho_i^c$ do not enter as mixed terms in $\rho$, thereby precluding alternative decompositions into $\rho_i^{\prime c}$. In summary, in the presence of noise, the chaos anticipated from the mean-field approximation is absent, whereas the Zeno effect can induce genuine chaos in the system.

\section{full quantum analysis}
\label{full quantum analysis}
\begin{figure*}[]
\centering
\begin{subfigure}{}
    \includegraphics[width=4in]{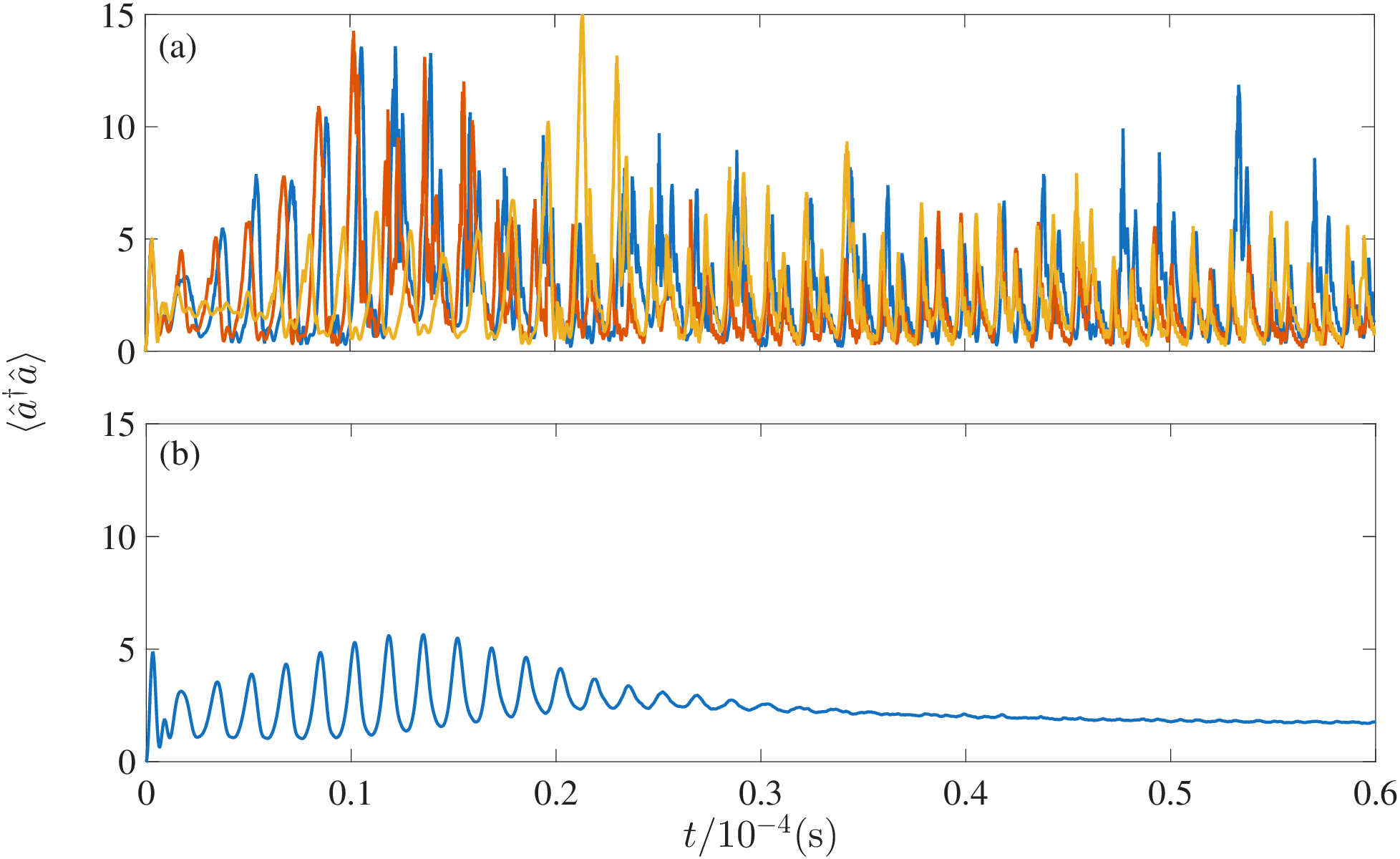}
    \label{fig:subfig1}
\end{subfigure}
\begin{subfigure}{}
    \includegraphics[width=4.5in]{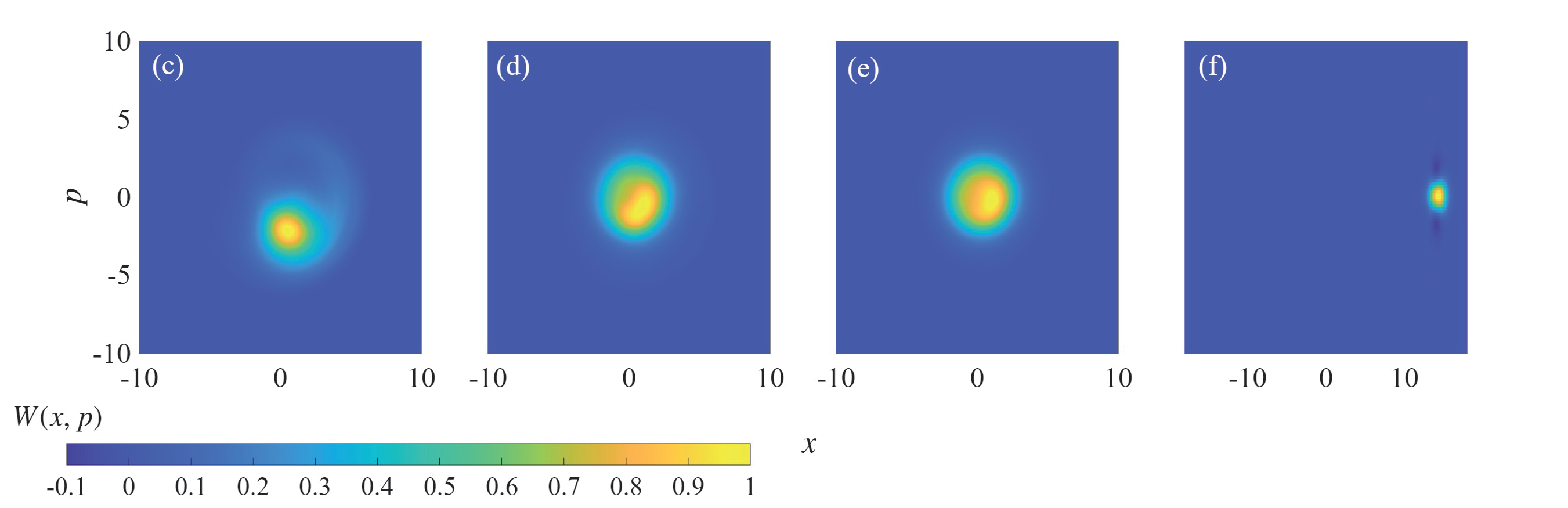}
    \label{fig:subfig2}
\end{subfigure}
\caption{(a): The expected value of the light intensity, $\langle\psi\vert \hat{a}^\dagger\hat{a}\vert\psi\rangle$, calculated from the quantum trajectories corresponding to three quantum jumps. (b): The expected value of the light intensity, $\text{Tr}( \hat{a}^\dagger\hat{a}\rho)$, calculated from the density matrix obtained from $5000$ trajectories. (c)-(f): Wigner functions of the cavity field state at times $t/10^{-4}=0.15$\,s, $t/10^{-4}=0.3$\,s, $t/10^{-4}=0.45$\,s, and $t/10^{-4}=0.6$\,s. Here, while keeping $C_1$ and $C_2$ unchanged, we set $g/2\pi=200$~kHz, i.e., $g/\omega_m= 0.38$. Other parameters are the same as in Fig.~\ref{fig:2}.}
\label{fig:5}
\end{figure*}

We have shown that fluctuations arising from thermal noise at room temperature can substantially suppress the onset of chaos, whereas cooling and noise mitigation can reinstate chaotic dynamics. In the idealized scenario devoid of fluctuations, the system's evolution reduces to that described by the mean-field approximation. However, this fluctuation-free limit pertains solely to classical mechanics; the Heisenberg uncertainty principle renders zero-point fluctuations unavoidable. Although these fluctuations are small in magnitude, it is noteworthy that Eqs.~\eqref{mean_value} and \eqref{eq:cle} permit the scaling $\alpha = g a$, $\beta = g b$~\cite{Li2023,Zhang2025}. For the stochastic Langevin equation, this rescaling results in
\begin{equation}
\begin{split}
\dot{\alpha}=&\left\{i\left[\Delta+2\text{Re}\left(\beta\right)+C_2\vert \alpha\vert^2\right]-\kappa\right\}{\alpha}+C_1+g\sqrt{2\kappa}{a}_{\rm in}, \\
\dot{{\beta}}=&(-i\omega_b-\gamma){\beta}+i\left(\vert a\vert^2-\dfrac{1}{2}\right)+g\sqrt{2\gamma}{b}_{\rm in},
\end{split}
\label{eq:cle_g}
\end{equation}
where $g E = C_1$ and $\chi/g^2 = C_2$ serve as renormalized dynamical parameters, with $g$ acting as a scaling factor. Consequently, with $C_j$ held constant, the mean-field dynamics of the system remain invariant, yet the effective noise amplitude increases with the nonlinear coupling strength $g$.
This implies that, for sufficiently strong nonlinearity, zero-point fluctuations are amplified to a degree capable of disrupting chaotic behavior, and such disruption is inevitable due to the inherent nature of Heisenberg-induced fluctuations. In essence, quantum mechanics intrinsically impedes the emergence of chaos in strongly nonlinear regimes. The mechanism described above remains at the level of semiclassical approximation, but it can be verified through a fully quantum analysis. To elucidate this, we maintain fixed values of  $C_1$ and $C_2$ while varying $g/\omega_b \in [10^{-1}, 10^0]$, thereby ensuring that the excitation numbers in modes $a$ and $b$ are on the order of $10^1$ to $10^2$. The complete quantum dynamics are then described by the master equation within a truncated Hilbert space~\cite{Liao2016}:
\begin{equation}
\begin{split}
\dfrac{d}{dt}\rho=-i[H,\rho]+\kappa\left(2\hat{a}\rho\hat{a}^\dagger-\hat{a}^\dagger\hat{a}\rho-\rho\hat{a}^\dagger\hat{a}\right)+\gamma\left(2\hat{b}\rho\hat{b}^\dagger-\hat{b}^\dagger\hat{b}\rho-\rho\hat{b}^\dagger\hat{b}\right),
\end{split}
\label{eq:master_eq}
\end{equation}
We employ the standard quantum jump method to simulate the evolution of this master equation. Specifically, for a quantum state $\vert\psi(t_0)\rangle$, after an evolution over a time interval $dt$, the final state will probabilistically be in one of three states:
\begin{equation}
\begin{split}
&\vert \psi_0(t_0+dt)\rangle=\dfrac{e^{-i\hat{H}_{\rm eff}dt/\hbar}\vert\psi(t_0)\rangle}{\sqrt{1-\sum_{j=1,2}P_j}},\\
&\vert \psi_1(t_0+dt)\rangle=\sqrt{\dfrac{2\kappa dt }{P_1}}\hat{a}\vert\psi(t_0)\rangle,\\
&\vert \psi_2(t_0+dt)\rangle=\sqrt{\dfrac{2\gamma dt }{P_2}}\hat{b}\vert\psi(t_0)\rangle,\\
\end{split}
\label{eq:jump_eq}
\end{equation}
where $\hat{H}_{\rm eff}$ is the non-Hermitian Hamiltonian $\hat{H}_{\rm eff}/\hbar=\hat{H}/\hbar-i\kappa \hat{a}^\dagger\hat{a}-i\gamma \hat{b}^\dagger\hat{b}$, and their respective probabilities are:
\begin{equation}
\begin{split}
&P_0=1-\sum_{j=1,2}P_j,\\
&P_1=2\kappa dt\langle \psi (t_0)\vert\hat{a}^\dagger\hat{a}\vert\psi(t_0)\rangle,\\
&P_2=2\gamma dt\langle \psi (t_0)\vert\hat{b}^\dagger\hat{b}\vert\psi(t_0)\rangle.\\
\end{split}
\label{eq:P}
\end{equation}
 This method significantly increases the dimension of the Hilbert space we can simulate. Specifically, the cavity field dimension is set to $\text{dim}_a=20$, and the oscillator dimension is set to $\text{dim}_b=600$, resulting in a total dimension of $\text{dim}_{\rm total}=12000$. In the quantum jump method, each calculation yields a trajectory of pure state evolution, and the system's density matrix is obtained from $N$ such calculations, that is, 
\begin{equation}
\begin{split}
\rho(t)=\dfrac{1}{N}\vert \psi(t)\rangle\langle \psi(t)\vert.
\end{split}
\label{eq:rho}
\end{equation}
After obtaining the density matrix, the expectation value of the light intensity is given by $\langle \hat{a}^\dagger \hat{a} \rangle = \text{Tr}(\hat{a}^\dagger \hat{a} \rho)$. The quantum state of the cavity field at this time corresponds to the reduced density matrix $\rho_a = \text{Tr}_b(\rho)$, which can be described by the Wigner function, whose expression is~\cite{book3}: 
\begin{equation}
\begin{split}
W(x,p)=&\dfrac{1}{\pi}\int_\infty^\infty dy\langle x-y\vert\rho\vert x+y\rangle e^{2ipy}\\&=\sum_{mn}\langle m\vert \rho\vert n\rangle W_{mn}(x,p),
\end{split}
\label{eq:Wigner}
\end{equation}
where $m$ and $n$ denote the indices of the Fock states, and we define
\begin{equation}
\begin{split}
W_{mn}(x,p)=&\dfrac{1}{\pi}\int_\infty^\infty dy\psi_m(x-y)\psi_n(x+y) e^{2ipy},\\
\end{split}
\label{eq:Wigner_1}
\end{equation}
where $\psi_n(x)$ is the Fock state $\vert n\rangle$ in the position basis:
\begin{equation}
\begin{split}
\psi_n(x)=\left( \dfrac{1}{\pi4^n(n!)^2} \right)^{\frac{1}{4}}\exp\left(-\dfrac{x^2}{2}\right)H_n(x).\\
\end{split}
\label{eq:Wigner_2}
\end{equation}
$H_n(x)$ is the Hermite polynomial of degree $n$. We plot the corresponding results in Fig.~\ref{fig:5}(b). As evident, within the full quantum dynamics, the light intensity continues to exhibit regular evolutionary characteristics, in contrast to the predictions of the mean-field approximation. The corresponding Wigner function, however, gradually deviates from a Gaussian state.

Figure~\ref{fig:5}(a) depicts the evolution of three randomly selected quantum trajectories. As evident, the computed evolution of the mechanical quantities for a single trajectory manifests irregular, chaotic behavior. Nonetheless, owing to the chaotic dynamics, the peaks and valleys of these trajectories are out of phase. We reiterate that, in the absence of continuous counting measurements on the system, a quantum trajectory associated with a single quantum jump serves merely as a mathematical auxiliary construct devoid of observable physical significance. Only the ensemble average of these trajectories, the density matrix in Eq.~\eqref{eq:rho},  possesses physical meaning, as alternative decompositions of the density matrix are always possible. As illustrated in Fig.~\ref{fig:5}(b), the dynamical evolution derived from the mixed state distinctly reverts to regular dissipative behavior, with its profile closely resembling that of the high-noise semi-classical case [the blue curve in Fig.~\ref{fig:3}(c)]. Given that the master equation~\eqref{eq:master_eq} accounts for a vacuum environment, devoid of thermal excitations in the reservoir, from a quantum mechanical standpoint, chaos in strongly nonlinear systems is inevitably suppressed by the Heisenberg uncertainty principle.

It is noteworthy that the distinction between the full quantum and semiclassical treatments emerges in the system's state following the suppression of chaos. In the semiclassical regime, after phase diffusion quenches chaos in the time domain, the probability distribution function of the system's state in phase space assumes the form of an attractor. In the full quantum simulation, during the initial stage [Fig.~\ref{fig:5}(c)], the Wigner function similarly exhibits a diffusion effect in phase space. However, with progressing time, it does not develop a chaotic configuration but instead manifests a distribution featuring negative values. This observation suggests that, in quantum mechanics, as nonlinearity intensifies, the limit cycle-to-chaos phase transition is supplanted by a limit cycle state-to-negative Wigner function state phase transition. Given that negative values in the Wigner function signify the quantum character of the system, this implies that, within the framework of nonlinear systems, the quantum analog of chaos is, in fact, quantumness itself.

\section{Discussion and conclusion}
\label{Discussion and conclusion}
We investigate the annealing of chaotic dynamics in nonlinear quantum-optical systems, focusing on how the chaotic trajectories predicted by mean-field equations transition to regular evolution of classical expectation values upon inclusion of fluctuations and noise. The underlying physical mechanism arises from a hallmark of chaos: sensitivity to weak initial perturbations from noise or fluctuations, which profoundly influence the subsequent dynamics and thereby regularize the expectation-value trajectories. We analyze the nonlinear dynamics of a prototypical optomechanical system in the semiclassical regime, characterized by strong driving and weak nonlinearity, by mapping the quantum Langevin equations to their stochastic counterparts. Our results indicate that irregularity persists marginally under low-noise conditions but is substantially suppressed, yielding regular evolution, at higher noise levels. Furthermore, we examine the phase-space distribution of the system state and observe that its form aligns closely with the phase-space trajectory predicted by mean-field theory. Consequently, for chaotic systems, the relaxation to an asymptotic steady state can be interpreted as a dephasing process, wherein undamped phase fluctuations lead to the distribution spreading over the entire classical orbit. While chaotic dynamics can be reinstated by reducing the ambient temperature, the ultimate limit of cooling, induced by zero-point fluctuations arising from the Heisenberg uncertainty principle, is invariably present. In strongly nonlinear systems, even these zero-point fluctuations can be amplified to a degree sufficient to suppress the onset of chaos. This conclusion has been corroborated through the quantum master equation and illustrated via the Wigner function of the corresponding quantum state. Furthermore, we have observed that, upon complete suppression of chaos, the semiclassical probability distribution of the system state shows a chaotic form, whereas the full quantum Wigner function retains a regular structure, albeit with the emergence of negative values. This indicates that, from a quantum mechanical viewpoint, the limit cycle-to-chaos phase transition is supplanted by a limit cycle-to-negative Wigner function phase transition; alternatively, we posit that the quantum analog of the chaotic state corresponds to the system manifesting enhanced quantum features. Given that negative Wigner functions lack a classical counterpart, we may further conclude that quantum mechanics inherently prohibits the emergence of chaos.

It is noteworthy that this physical mechanism constitutes a universal phenomenon. Indeed, our examination of chaotic systems reported in Refs.~\cite{Peng2024,Saikoa2025} yields consistent results. Furthermore, we present an analysis demonstrating that, when considering the system's dynamics via a textbook-style quantum measurement, wherein the system evolves to time $t$ before a single measurement is performed, with expectation values and results at other times obtained through repeated experiments, chaotic dynamics are absent in this case. In contrast, continuous weak probing of the system induces the Zeno effect, thereby restoring chaotic behavior. Our findings offer an intuitive framework for understanding the classical-quantum crossover in nonlinear systems.
\begin{acknowledgements}
W.~L. is supported by the National Natural Science Foundation of China (Grants No.~12304389), the Scientific Research Foundation of NEU (Grant No. 01270021920501*115). X.~L. and J.~C. are supported by the Research Foundation of National Key Laboratory of Shock Wave and Detonation Physics (Grant No. 2024CXPTGFJJ06407). X.-l. Li is supported by the National Natural Science Foundation of China (Grant No. 12247101), the Fundamental Research Funds for the Central Universities (Grant No. lzujbky-2025-jdzx07), the Natural Science Foundation of Gansu Province (No.25JRRA799), and the ‘111 Center’ under Grant No. B20063.
\end{acknowledgements}

\end{document}